\documentclass[conference]{IEEEtran}
\IEEEoverridecommandlockouts
\usepackage{amsmath,amssymb,amsfonts}
\usepackage{listings}
\usepackage{algorithm2e}
\usepackage{graphicx}
\usepackage{textcomp}
\usepackage{xcolor}
\usepackage{tikz}
\usetikzlibrary{shapes,positioning,calc}
\usepackage[hyphens]{url}
\usepackage[unicode, hidelinks]{hyperref}
\usepackage[hyphenbreaks]{breakurl}
\usepackage[backend=biber, style=ieee, bibencoding=utf8, citestyle=numeric-comp]{biblatex}
\def\BibTeX{{\rm B\kern-.05em{\sc i\kern-.025em b}\kern-.08em
    T\kern-.1667em\lower.7ex\hbox{E}\kern-.125emX}}
\usepackage{booktabs}
\usepackage{colortbl}
\usepackage{multirow}
\usepackage{flushend}
\usepackage{pgfplots}
\begin{document}

\title{Towards Symbolic Pointers Reasoning in Dynamic Symbolic Execution
\thanks{This work was supported by RFBR grant 20-07-00921 A.}}

\author{
\IEEEauthorblockN{
  Daniil Kuts
}
\IEEEauthorblockA{
  Ivannikov Institute for System Programming of the RAS
}
Moscow, Russia \\
kutz@ispras.ru
}

\maketitle

\begin{tikzpicture}[remember picture, overlay]
\node at ($(current page.south) + (0,0.65in)$) {
\begin{minipage}{\textwidth} \footnotesize
  Kuts D. Towards Symbolic Pointers Reasoning in Dynamic Symbolic Execution.
  2021 Ivannikov Memorial Workshop (IVMEM), IEEE, 2021, pp. 42-49, DOI:
  \href{https://www.doi.org/10.1109/IVMEM53963.2021.00014}{10.1109/IVMEM53963.2021.00014}.

  \copyright~2021 IEEE. Personal use of this material is permitted. Permission
  from IEEE must be obtained for all other uses, in any current or future media,
  including reprinting/republishing this material for advertising or promotional
  purposes, creating new collective works, for resale or redistribution to
  servers or lists, or reuse of any copyrighted component of this work in other
  works.
\end{minipage}
};
\end{tikzpicture}

\begin{abstract}
Dynamic symbolic execution is a widely used technique for automated
software testing, designed for execution paths exploration and
program errors detection. A hybrid approach has recently become widespread,
when the main goal of symbolic execution is helping fuzzer increase program coverage.
The more branches symbolic executor can invert, the more useful it is for
fuzzer. A program control flow often depends on memory values, which are obtained
by computing address indexes from user input. However, most DSE tools
don't support such dependencies, so they miss some desired program branches.

We implement symbolic addresses reasoning on memory reads in our dynamic symbolic
execution tool Sydr. Possible memory access regions are determined by either
analyzing memory address symbolic expressions, or binary searching with SMT-solver.
We propose an enhanced linearization technique to model memory accesses.

Different memory modeling methods are compared on the set of programs. Our evaluation
shows that symbolic addresses handling allows to discover new symbolic branches and
increase the program coverage.

\end{abstract}


\section{Introduction}

Automated testing tools allow developers to find program errors in advance and improve
the software quality. The state-of-the-art dynamic testing tools are based
on the two main technologies -- coverage-based fuzzing~\cite{sargsyan19, fioraldi20, aschermann19}
and dynamic symbolic execution~\cite{poeplau20, vishnyakov20, cha12, saudel15, yun18}. Recently,
a hybrid approach~\cite{stephens16, borzacchiello21, poeplau21}, when fuzzer and symbolic
executor are run side by side, has proven itself as the most effective way to fuzz binary
programs. A fuzzer, with fast and lightweight mutations, can quickly discover
new coverage. Besides, a symbolic executor, performing slow and complex but more accurate
analysis, can invert difficult branches and detect bugs on already discovered paths.

Most symbolic execution implementations consider only direct transferring of symbolic
data between instruction source and destination operands, which leads to missing certain
branches. Let's consider the program code example with  table dependency:
\begin{lstlisting}[language=C, basicstyle=\small\ttfamily, numbers=left,
xleftmargin=2em, breaklines=true]
int table[7] = {3, 7, 14, 0, 5, 11, 9};
int foo(int a) {
    int res = table[a];
    if (res == 5) {
        abort();
    }
    return res;
}
\end{lstlisting}
Depending on the input data, an element is read from the array, which then used in
target branch. The following assembly corresponds to the given code:
\begin{small}
\begin{verbatim}
780: lea    rax,[rip+0x200899] # 201020 <table>
787: movsxd rdi,edi
78a: mov    eax,DWORD PTR [rax+rdi*4]
78d: cmp    eax,0x5
790: je     794 <foo+0x14>
792: repz ret
794: sub    rsp,0x8
798: call   5b0 <abort@plt>
\end{verbatim}
\end{small}
A base address of the table is loaded in the line 1. It is used to compute an address of
the certain element in line 3, which then used in comparison operation. Normally, the result
of memory read in line 3 would be concretized, thus, missing the target branch in line 5.
Described indirect data dependencies are common in the programs. For instance, standard
\texttt{tolower}/\texttt{toupper} functions or hashing algorithms use transformation tables.

Symbolic execution tools have different methods for handling symbolic addresses~\cite{baldoni18}.
Mayhem~\cite{cha12} utilizes an index-based memory model and processes
symbolic reads by building a binary search tree over possible address ranges, that are
determined by using SMT-solver and value set analysis~\cite{reps04}. Mayhem proposes a
linearization method for optimizing a binary search tree depth by merging leafs with a single
linear formula. Different approach is used in KLEE~\cite{cadar08}, a symbolic execution engine
based on LLVM IR. Instead of using flat memory model, it tracks active memory objects during
symbolic execution. On every load or store at a symbolic address KLEE queries SMT-solver
to determine a list of objects that can be referred by this access. Then it forks symbolic
execution for every found memory object while constraining symbolic addresses according to memory
bounds. The state-of-the-art concolic executor QSYM~\cite{yun18} handles symbolic addresses
by simply fuzzing them. When QSYM encounters symbolic address, it starts repeatedly
querying SMT-solver to get maximum and minimum address values and produces new input on
every solver invocation.

We implemented symbolic addresses reasoning in Sydr~\cite{vishnyakov20}. The developed
approach is based on methods proposed by Mayhem and works for the analysis of x86 binary code.
Every time Sydr encounters a memory read we check whether it have a symbolic address and try
to determine approximate memory bounds. Finally, using improved linearization technique,
we build an expression, that models a memory read according to a symbolic address.

This paper makes the following contributions:
\begin{itemize}
  \item We implement the symbolic addresses processing with different
  techniques: building a nested if-then-else tree, a binary search tree,
  and linearization approach. We evaluate the effectiveness of each method on
  the set of real-world programs. Generated queries are processed by different
  SMT-solvers.
  \item We propose an improvement of the linearization approach, that allows us
  to reduce solving time for generated SMT-queries. Moreover, the proposed method
  handles not only symbolic addresses, but also symbolic memory values that may
  appear at these addresses.
\end{itemize}

The rest of this paper is organized as follows. Section~\ref{symaddr_reasoning}
describes symbolic addresses handling. Section~\ref{implementation} explains how
it is implemented in Sydr. The experimental evaluation is given in
Section~\ref{evaluation}. Finally, Section~\ref{conclusion} concludes this paper.

\section{Symbolic Addresses Reasoning}
\label{symaddr_reasoning}

Processing of memory access at a symbolic address means modeling simultaneous
accesses to multiple memory cells, depending on the values that a symbolic address
can hold~\cite{coppa17}. Memory can be accessed both on the read and on the write.
To model a memory read at a symbolic address it is enough to describe the contents
of memory cells that can be accessed in a single expression. That expression is used to
build the instruction semantics, which then assigned to one destination register.
Modeling a memory write at a symbolic address requires to assign an instruction result
expression to every memory cell, that can be accessed by address. Besides, for each memory
cell the expression should be additionally constrained with specific symbolic address precondition.
Unlike symbolic reads, this time a whole memory region becomes symbolized instead of one register.
That leads to an exponential growth of the tainted data, which, in turn, negatively affects the
symbolic execution performance. We process only symbolic memory reads in order to keep analysis
scalability at an acceptable level.

A switch table~\cite{cifuentes01} is a specific case of an address dependency. Unlike the regular
memory accesses, we know exactly where the branching instruction is placed. So, in this case we
build path constraints for each possible jump target instead of modeling a memory access~\cite{vishnyakov20}.

\subsection{Boundaries Approximation}

One of the problems in symbolic addresses reasoning is determining the range
of values that a symbolic address can hold. It is essential to find the bounds
that are closest to the real ones, since the underestimation of address range leads
to inaccurate symbolic execution and possible losses of execution paths. On the
contrary, an enormous address range increases the size of symbolic expression for memory
access. This leads to a higher memory consumption, increased SMT solver workload, and,
as a result, to the analysis performance drop.

For specific cases we could guess address bounds by analyzing the accessed
memory values. For instance, the mentioned above jump tables are restricted to contain
either valid pointers to executable code or offset values used to compute such code
pointers. For this case determining memory access bounds is quite easy and
the only problem is distinguishing adjacent jump tables. But in general case the memory
contents are not unified in any way and it is not possible to make any assumptions.
The basic way to determine memory bounds is selecting a memory region of some
constant length. Boundaries are selected at an equal distance in both directions
from concrete address value, so that the entire memory region holds a given number
of elements, according to the size of memory access. The length is measured in the
number of elements in order to parse memory for symbolic reads of different sizes
equally. This method has extremely low accuracy, so we use it only when all other
methods have failed.

We use some heuristics to deduce the lower bound of memory access from the
symbolic address expression. Instructions that perform random memory access
have the address encoded in several components. For instance, the address for
\texttt{WORD PTR[rax + rbx*1 + 0x100]} memory access would be calculated as
$base + index * scale + displacement$, where \texttt{rax} is a base register and
\texttt{rbx} is index. If address is symbolized, then some of its components
are symbolic and the whole value may vary within the range of symbolic part against
concrete part. The main idea behind this approach is that, generally, the concrete
part is the base address of some table in program memory and symbolic part is an
offset in this table, that may vary according to user input. By analyzing
an abstract syntax tree (AST) of address expression we can extract its concrete part
and assume that it is a base address of memory access, i.e. the lower bound. The
main shortcoming of this approach is that concrete part of expression often contains
constant index or displacement. In this case we traverse the concrete part AST
to decompose it onto terms. The largest term is then assumed to be the base address.
Also this approach can't be applied to find the address upper bound, so another method
should be used for that. The described approach has average accuracy and helps to reduce
the possible memory range in many cases, but still may fail to detect base address due
to complex expression AST or non-typical address computation. If base address doesn't
seem to be correct, we choose a left border at a constant length from current access address.

The most accurate but the most expensive way to find memory bounds is utilizing
SMT-solver. We can determine maximum and minimum address values by slicing~\cite{vishnyakov20}
path predicate based on symbolic address expression and repeatedly querying solver.
First, we assume that symbolic address bound may lay between its concrete
value and certain limit value, determined by maximum memory length. Then we perform
the binary search to determine the exact boundary value. If solver decides
that symbolic address can exceed the limit value, then the limit value is set as
boundary. Due to symbolic execution shortcomings, the symbolic address may be overconstrained
to its concrete value, that is, it cannot take any other value. Still,
such symbolic reads should be processed, as it could become changeable in optimistic
solving~\cite{yun18}. Using multiple solver invocations on every symbolized memory read causes
a huge performance drop even more severe than the one caused by excessive memory
range length. So, this technique is applicable only on simple programs and where
knowing exact memory access bounds is crucial.

Also there is one more method to find length of symbolic address range.
Sometimes programs can validate the symbolic memory access address in preceding
basic blocks. A constant value in a comparison instruction with symbolic part of
address presumably is a memory range length. This approach was not implemented but
planned as a future work.

To summarize, we always use symbolic AST analysis for the
lower bound reasoning. SMT-solver or constant length are utilized when AST analysis
fails. Also these two methods are used to determine the address upper bound. We use
SMT-solving only in extreme cases, when high accuracy is required or the analyzed
program is small enough and produces formulas that are easy to solve.

\subsection{Modeling Memory Accesses}
\label{model_memory}

After the memory access bounds are determined, a formula that models a symbolic
read from this memory region should be built. Given a set of concrete addresses
and corresponding memory values, a formula should define how the operation result
depends on the symbolic address. The simplest method to model symbolic read is
iterating over all possible memory values and building if-then-else (ITE) tree:

\SetAlgoSkip{}
\begin{algorithm}
\DontPrintSemicolon
\small
$sym \gets symbolic\_address$\;
\lIf {$sym == a1$} {
    $value\_1$
}
\Else {
    \lIf {$sym == a2 \lor sym == a3$} {
        $value\_2$
    }
    \Else {
        \lIf {$sym == a4$} {
            $value\_4$
        }
        \lElse {
            $current\_value$
        }
    }
}
\end{algorithm}
%\begin{small}
%\begin{verbatim}
%sym = symbolic_address
%if (sym == a1) then value_1
%else
%    if (sym == a2 || sym == a3) then value_2
%    else
%        if (sym == a4) then value_4
%        else cur_value
%\end{verbatim}
%\end{small}

For better understanding, the above algorithm demonstrates the semantics of an actual formula, that is built
in SMT-LIBv2 language~\cite{smt}. Additionally, we combine memory cells with the
same values into one node to reduce the tree nesting level. Because of inaccurate
memory bounds identification, sometimes symbolic address actually can go beyond the
selected memory region. It is possible to constrain symbolic address expression
to be within the assumed bounds, but additional assertions would produce larger
path predicates and, moreover, would overconstrain symbolic execution model,
thereby pruning potential execution paths. A better way may be to tie all
unexpected address values to the memory value on current execution path. Although
such modeling is incomplete, it doesn't constrain the path predicate feasibility
and sometimes allows to produce inputs which have a chance to explore new execution
paths. This approach doesn't require a separate assertion and can be implemented
simply as a last else-node of the tree.

Another technique of a symbolic read modeling is constructing a binary search
tree (BST) over the range of possible address values. Symbolic address expression
is recursively compared with subranges, leafs of the tree are the corresponding
memory values. BST have less depth than its nested analogue, which reduces a solver
workload. We handle addresses outside the selected memory region similarly to the ITE tree.
There are additional leafs in BST with current memory value for such addresses:

\SetAlgoSkip{}
\begin{algorithm}
\DontPrintSemicolon
\small
$sym \gets symbolic\_address$\;
\If {$sym < 0x300$} {
    \lIf {$sym < 0x100$} {
        $current\_value$
    }
    \Else {
        \lIf {$sym == 0x100$} {
            $value\_1$
        }
        \lElse {
            $value\_2$
        }
    }
}
\Else {
    \lIf {$sym < 0x400$} {
        $value\_2$
    }
    \Else {
        \lIf {$sym == 0x400$} {
            $value\_3$
        }
        \lElse {
            $current\_value$
        }
    }
}
\end{algorithm}
%\begin{small}
%\begin{verbatim}
%sym = symbolic_address
%if (sym < 0x300)
%then
%    if (sym < 0x100) then cur_value
%    else
%        if (sym == 0x100) value_1
%        else value_2
%else
%    if (sym < 0x400) then value_2
%    else
%        if (sym == 0x400) then value_3
%        else cur_value
%\end{verbatim}
%\end{small}

\begin{figure}
\begin{tikzpicture}
\begin{axis}[
  xlabel={Byte array indexes},
  ylabel={Memory values},
  xtick={0,4,8,12,16,20,24,28,32},
  ytick={1,9,13,15,17},
  ymajorgrids=true,
  grid style=dashed,
  xlabel near ticks,
  ylabel near ticks,
  width=\textwidth/2,
  height=\textheight/3,
]

\addplot[
  color=black,
  mark=*,
  ]
  coordinates {
  (0,1) (4,9) (8,17)
  }node[pos=0.7,anchor=east,rotate=68, yshift=6]{$y=2x+1$};

\addplot[
  color=black,
  mark=*,
  ]
  coordinates {
  (16,13) (20,1)
  }node[pos=0.2,anchor=west,rotate=282, yshift=7]{$y=-3x+61$};

\addplot[
  color=black,
  mark=square,
  ]
  coordinates {
  (12,15) (28,15)
  }node[pos=0.3,anchor=south, yshift=0]{$y=15$};

\addplot[
  color=black,
  mark=square,
  ]
  coordinates {
  (24,17)
  }node[pos=0.5,anchor=west, yshift=3]{$(24, 17)$};
\end{axis}
\end{tikzpicture}
\caption{Merging memory points with linear functions.}
\label{fig:lin}
\end{figure}
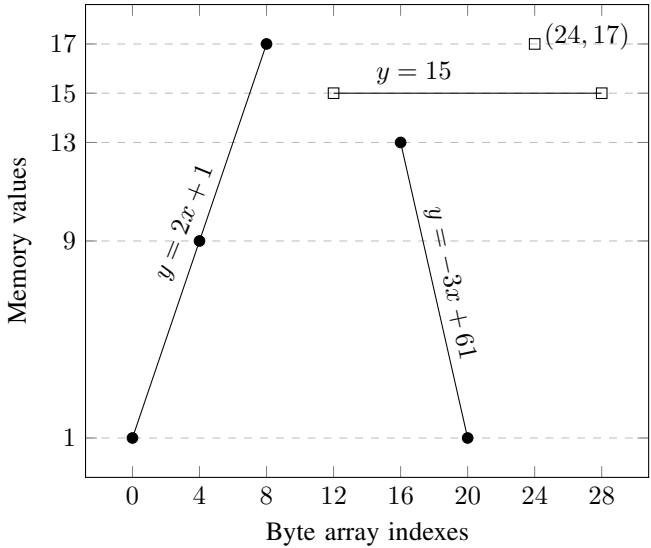

Described BST can be optimized by merging multiple leafs with a
single linear formula. This approach is proposed by Mayhem~\cite{cha12} and is called
linearization. Memory region is represented as a set of points on the plot, where
the axes correspond to memory values and indexes. The original linearization algorithm~\cite{cha12}
suggests using indexes instead of addresses. That is, all memory points on the plot should be
incremented by one on the abscissa: 0, 1, 2, etc. In our approach we have symbolic expression
only for the whole address and can't properly extract index expression in general. Thus,
we use address values as indexes to represent memory points on the plot. We normalize address
values relatively to the lower bound, because concrete addresses are huge enough. Therefore,
we have small values on the abscissa, which are incremented by memory access byte size.
Figure~\ref{fig:lin} shows how memory region with 8 entries and 4 byte memory access may be
presented on the plot. Of course, it is possible to convert addresses to the indexes in our
approach too. By dividing normalized address values by the memory access size we can get
Mayhem-like indexes. But our experiments have shown that this option has much worse results,
mostly due to the division operations in formulas. Further, we draw lines through the memory
points so that the following conditions are met:
\begin{itemize}
  \item Line equations should contain only integer coefficients to be represented
  in bitvector integer logic. Our indexes are not always incremented by one. Thus,
  despite   a line can be drawn through any two points, its coefficients may be fractional.
  \item Lines should be drawn only through the sequential points: no point should be
  skipped when drawing a line. This condition obliges all lines to be ordered,
  so we could easily build a BST over them.
\end{itemize}

We iterate over all points and compute a line through two adjacent points.
If this line have integer coefficients, then we start checking whether next points
lay on it and stop on the first point that doesn't. We save that line, remove
all joined points and continue algorithm from the point we stopped checking on.
As a result, we have a set of lines and remaining points, which don’t belong to
any line. To reduce the number of single points we check whether any of them have
the same memory value so that they can be joined with a single horizontal
line. Because this line doesn't go through the adjacent points it only could be
appended as an extra ITE node before the BST. However, this still helps to reduce the
final tree depth.

Computed linear functions are represented by bitvectors of a certain size, that
should be able to fit both memory values and indexes. We try to pick a minimal
possible size, so that line coefficients and a final expression would be more
compact. We can't reduce size for memory value nodes, but can do it for index nodes. They have
original size of symbolic address node, but contain much smaller values because of
normalization. We could try to reduce the index node size if memory access size is smaller.
In the same time, index node size should be able to contain largest index value for memory bounds.
The equation bitvector size can't be smaller than memory access size. So, in case of memory access
of a large size (128-bit or larger), the expressions for linear functions are too big. As a workaround
we switch from linearization approach to a nested ITE tree building for such memory reads.

It often appears that besides the symbolic memory access address, the memory values
are symbolized too. The symbolic expression of memory cell should be
taken instead of its concrete value. During the construction of simple nested tree
or BST the corresponding memory value node is replaced with the required symbolic
node. However, symbolized memory can’t be processed by linearization approach because
it requires concrete values to compute a line. Concretizing memory values may help
to build effective linearized BST, but symbolic execution accuracy will suffer.
Thus, we treat symbolic and concrete memory values separately. We use a linearization approach
for concrete values, while symbolic memory is handled by building a classic nested tree.
This nested tree then prepended before the linearized BST just as in the case of horizontal lines.
The linearized BST for Figure~\ref{fig:lin} is stated below:

\SetAlgoSkip{}
\begin{algorithm}
\DontPrintSemicolon
\small
$sym \gets symbolic\_address - base\_address$\;
\lIf {$sym == 12 \lor sym == 28$} {
    $15$
}
\Else {
    \If {$sym < 24$} {
        \lIf {$sym < 16$} {
            $2 * sym + 1$
        }
        \lElse {
            $-3 * sym + 61$
        }
    }
    \Else {
       \lIf {$sym < 32$} {
            $17$
        }
        \lElse {
            $current\_value$
        }
    }
}
\end{algorithm}
%\begin{small}
%\begin{verbatim}
%sym = symbolic_address - base_address
%if (sym == 12 || sym == 28) then 15
%else
%    if (sym < 24)
%    then
%        if (sym < 16) then 2*sym+1
%        else -3*sym+61
%    else
%        if (sym < 32) then 17
%        else cur_value
%\end{verbatim}
%\end{small}

Unlike the default BST approach, linearization formula requires only upper bound
checking to constrain symbolic address to the selected memory region. When address
value goes outside the lower bound, the symbolic index expression used in formula
wraps around due to bitvector integer logic. Thus, it becomes too huge and falls
beyond the upper bound.

\section{Implementation}
\label{implementation}

The proposed method for symbolic addresses reasoning is implemented in the
dynamic symbolic execution tool Sydr~\cite{vishnyakov20}. Sydr is divided in two
separate processes -- Concrete Executor based on DynamoRIO framework~\cite{bruening04}
and Symbolic Executor based on Triton engine~\cite{saudel15}. Concrete Executor
fetches an instruction to be executed next, disassembles it, and sends all required
data (opcode, operands, concrete values, etc.) to symbolic executor via shared memory.
Symbolic Executor checks whether instruction operates on symbolized data and
depending on that either symbolically interprets it, or concretizes destination
operands and skips instruction processing.

Handling memory reads from symbolic addresses additionally requires transmission
of the entire memory region which will be modeled. Besides the fact that memory
load instructions are extremely common, the memory region parsing and transmission
are expensive operations. Instructions with memory reads are processed in two steps
to prevent unnecessary workload from non-symbolic memory accesses. First step is
identical to a normal instruction processing -- all information except memory is
passed to symbolic executor, which decides whether address of the memory access
operand is symbolized or not. If it is concrete, then instruction processing continues
as usual, otherwise the second step of processing begins. Symbolic executor tries to
determine the memory access bounds using methods described in Section 2.1. The estimated
memory bounds then passed back to Concrete Executor, which validates them and returns
corresponding memory values. Finally, Symbolic Executor uses these values to build a memory
read expression and interpret current instruction.

In Sydr, symbolic execution of the program is performed with Triton framework.
Normally, Triton concretizes all memory addresses and then loads (symbolic or concrete)
values from corresponding memory cells. Instead of changing Triton internals we
utilize its symbolic memory model: after building a memory read expression in a
way, described in Section~\ref{model_memory}, we assign it to the memory cell at concretized
address. Thus, we make this cell symbolized with our expression that correctly
describes a variable access to whole memory region. However, it is valid only
for the current moment of execution, so we restore the original state of memory
cell after Triton interprets an instruction.

\section{Evaluation}
\label{evaluation}

Due to symbolic address reasoning we are able to analyze many new program dependencies.
The overall symbolic execution speed has dropped, because now we symbolically process much more
instructions than before. In addition, each symbolized memory read is described by a large
tree-expression that covers a part of memory. Therefore, the symbolic formula sizes have
increased significantly, and so the memory consumption has increased. On the other hand, we
perform more complete program analysis and are able to discover new execution paths by
inverting symbolic branches, which previously were considered as independent from user input.
All our experiments were performed on the x86\_64 machine with two AMD EPYC 7702 64-Core processors
and 256Gb RAM. SMT-solver wasn't used for the memory bounds reasoning during evaluation, as it caused a
huge performance drop when analyzing real-world applications.

\begin{table}[ht]
\caption{SMT Expressions Solving Time}
\begin{center}
\scriptsize
\begin{tabular}{l @{\hspace{0.5\tabcolsep}} r @{\hspace{1\tabcolsep}} r @{\hspace{1\tabcolsep}} r | @{\hspace{1\tabcolsep}} r @{\hspace{1\tabcolsep}} r @{\hspace{0.8\tabcolsep}} r | @{\hspace{1\tabcolsep}} r @{\hspace{1\tabcolsep}} r @{\hspace{1\tabcolsep}} r}
\toprule
    \multirow{2}{*}{\textbf{Application}}&\multicolumn{3}{c}{\textbf{Z3}}&\multicolumn{3}{c}{\textbf{Yices2}}&\multicolumn{3}{c}{\textbf{Bitwuzla}} \\
    &\textbf{LIN}&\textbf{ITE}&\textbf{BST}&\textbf{LIN}&\textbf{ITE}&\textbf{BST}&\textbf{LIN}&\textbf{ITE}&\textbf{BST} \\
    cjpeg&\textbf{1m7s}&1m16s&1m14s&\textbf{1.2s}&2s&2.6s&30.8s&\textbf{7s}&20.6s \\
    eperl&\textbf{11.4s}&12.4s&20.3s&4.8s&\textbf{2.4s}&2.9s&17s&19.5s&\textbf{13.5s} \\
    foo2lava&\textbf{41.6s}&\textbf{42.5s}&\textbf{41.8s}&\textbf{1.4s}&\textbf{1.4s}&\textbf{1.5s}&\textbf{6.3s}&\textbf{6.5s}&\textbf{6.4s} \\
    hdp&\textbf{18s}&27.5s&19.9s&\textbf{1.7s}&2.1s&2.5s&\textbf{8.2s}&17s&13s \\
    jasper&\textbf{8.2}s&8.8s&9.8s&\textbf{1.4s}&\textbf{1.5s}&\textbf{1.5s}&\textbf{5.1s}&5.5s&5.7s \\
    libcbor\_cb&\textbf{11.5s}&T/O&17.4s&\textbf{0.8s}&1s&1.3s&\textbf{2.8s}&6.4s&4.3s \\
    libcbor\_map&\textbf{2.1s}&24.2s&4.5s&\textbf{0.3s}&0.6s&0.7s&\textbf{1s}&2.7s&2.5s \\
    libxml2&\textbf{4.2s}&\textbf{4.4s}&5.1s&\textbf{0.9s}&1.3s&1.5s&\textbf{3s}&8.8s&7.3s \\
    minigzip&\textbf{0.9s}&4.4s&8.6s&\textbf{0.2s}&1.4s&1.9s&\textbf{0.5s}&10.1s&9s \\
    muraster&\textbf{4.2s}&4.6s&5.8s&1.2s&\textbf{1s}&1.2s&\textbf{4.7s}&5.3s&5s \\
    openssl\_asn&13.5s&\textbf{3.6s}&4.7s&1.5s&\textbf{0.8s}&2s&6.2s&7.1s&\textbf{5s} \\
    openssl\_num&\textbf{2.4}s&6s&7.7s&\textbf{3s}&4.2s&5s&24s&\textbf{22s}&31s \\
    openssl\_x509&\textbf{2.5s}&3.1s&3.3s&\textbf{1s}&\textbf{0.9s}&1.2s&6.4s&10s&\textbf{5.5s} \\
    pk2bm&\textbf{5.2s}&\textbf{5.2s}&7s&\textbf{1s}&1.2s&1.5s&\textbf{2.6s}&6.8s&6.1s \\
    re2&\textbf{6.4s}&7s&7.2s&2.8s&\textbf{0.9s}&1.3s&\textbf{2.8s}&3.6s&5s \\
    readelf&\textbf{2.3s}&3.2s&3.6s&\textbf{1s}&\textbf{1.1s}&1.2s&13.4s&\textbf{11.5s}&\textbf{11.6s} \\
    sqlite3&\textbf{41s}&50s&47.6s&\textbf{3.7s}&5.5s&5.8s&19s&37s&\textbf{14.5s} \\
    suricata&\textbf{3.2s}&\textbf{3.2s}&\textbf{3.5s}&\textbf{0.6s}&\textbf{0.6s}&0.8s&\textbf{3.7s}&4s&4.4s \\
    yices&\textbf{5.1s}&\textbf{5.3s}&\textbf{5s}&0.9s&\textbf{0.7s}&0.9s&\textbf{3.2s}&4.4s&4s \\
    yodl&\textbf{4.7s}&6.9s&10s&\textbf{1.4s}&2.2s&2.4s&\textbf{3.9s}&9.5s&9.5s \\
    tiff2pdf&1m33s&\textbf{39s}&1m5s&8.6s&7.2s&\textbf{6.9s}&37.5s&\textbf{18s}&20s \\
\bottomrule
\end{tabular}
\label{tbl:smt}
\end{center}
\end{table}

Table~\ref{tbl:smt} shows the evaluation results for different memory modeling methods.
We implemented memory modeling with three methods: a linearization approach (column \texttt{LIN}),
a nested if-then-else tree (column \texttt{ITE}), and a binary search tree (column \texttt{BST}).
We enhanced linearization approach with additional horizontal lines and nested ITE tree for
symbolized memory values. Moreover, we switch to full nested ITE method for the large memory accesses.
So, column \texttt{LIN} displays results for our enhanced method. We ran our tool Sydr with different
methods enabled on the set of programs and collected SMT-queries to evaluate which of these approaches
are most suitable for solving. For each program we selected from dozens to hundreds of queries of different
size which contain many expressions for symbolic memory reads.
We choose three different solvers to process our queries: Z3~\cite{demoura08}, Yices2~\cite{dutertre14},
and Bitwuzla~\cite{niemetz20} to exclude individual features of SMT-solvers. The total
solution time for a set of queries was calculated and averaged over several runs. We use default solver
preferences without any optimization techniques.  The best solution time among different memory models
is highlighted in the Table~\ref{tbl:smt}. The results show that linearization
approach have the lowest decision time for the most of tested programs. Although, there are
several programs (\texttt{eperl}, \texttt{openssl\_asn1}, and \texttt{tiff2pdf}) where linearization
have the slowest solution time on the most solvers. All methods have close results on Yices2 solver. The
nested ITE approach is better than BST on Z3 solver and vice-versa on Bitwuzla.  Also Bitwuzla have the lowest
number of programs with linearization as the fastest method among other solvers. The nested ITE
queries for \texttt{libcbor\_cb} program were timed out when processing with Z3 solver as it exceeded 2 hours limit.
Generally, the linearization approach has either the best solution time or at least not worse,
except for a few programs. Also, switching to a nested ITE tree on large memory accesses helps us to process
some queries more effectively. So, in further experiments we used our enhanced linearization technique
as the memory modeling method.

%\begin{table*}[htbp]
\begin{table*}[t]
\caption{Analysis Efficiency}
\begin{center}
\scriptsize
\begin{tabular}{l @{\hspace{0.9\tabcolsep}} r >{\columncolor[gray]{0.9}}r r >{\columncolor[gray]{0.9}}r r >{\columncolor[gray]{0.9}}r r >{\columncolor[gray]{0.9}}r r >{\columncolor[gray]{0.9}}r}
\toprule
    \multirow{2}{*}{\textbf{Application}}&\multicolumn{2}{c}{\textbf{Path predicate time}}&\multicolumn{2}{c}{\textbf{Total time}}&\multicolumn{2}{c}{\textbf{Queries / min}}&\multicolumn{2}{c}{\textbf{SAT}}&\multicolumn{2}{c}{\textbf{Accuracy}} \\
    &\textbf{default}&\textbf{symaddr}&\textbf{default}&\textbf{symaddr}&\textbf{default}&\textbf{symaddr}&\textbf{default}&\textbf{symaddr}&\textbf{default}&\textbf{symaddr} \\
    cjpeg&18s&1m31s&60m&60m&5.3&5.1&56&54&89.3\%&92.6\% \\
    libxml2&15s&16s&9m59&60m&924.1&122.4&1247&1244&82.4\%&90.1\% \\
    readelf&27s&36s&60m&60m&85.7&13.1&2029&287&86.9\%&81.2\% \\
    libcbor&1.8s&2.1s&12s&1m58s&2176.5&210.2&275&295&100\%&40.6\% \\
    openssl&1m19s&1m38s&60m&60m&44.7&18.5&1000&234&75.7\%&70.5\% \\
    sqlite3&9.1s&10.7s&12m49s&14m56s&2871.5&2608.1&8414&10340&99.9\%&100\% \\
    minigzip&59s&3m48s&16m23s&60m&582.9&7.6&7569&238&51.5\%&100\% \\
    hdp&23s&31s&60m&60m&156.2&31.9&4417&962&73.7\%&68.3\% \\
    yices-smt2&10s&24s&22m22s&60m&494.1&50.6&5536&621&70.2\%&89\% \\
    yodl&6s&7s&9m2s&20m8s&852.3&396.1&1150&1421&98.3\%&98.3\% \\
	jasper&10m12s&16m16s&60m&60m&203&115.3&4164&3336&82.6\%&81.4\% \\
\bottomrule
\end{tabular}
\label{tbl:analysis}
\end{center}
\end{table*}

We analyzed the set of real-world applications with default symbolic execution and with symbolic addresses
reasoning enabled in order to evaluate the effect of supporting symbolic reads. Symbolic execution in Sydr
is performed in two steps. Firstly, the program is symbolically executed and its path predicate is built.
Secondly, SMT-solver starts inverting found branches. Table~\ref{tbl:analysis} allows to
evaluate the analysis performance drop from symbolic reads processing. The first column shows how addresses
processing increases the time of program symbolic emulation. The time could increase up to several times
depending on program and number of address dependencies it uses. The total analysis time was limited to 1 hour.
The second column shows that default symbolic execution has managed to invert all branches on more programs
within the time limit. That is, the speed of SMT-queries solving is higher for the default analysis.
The third column presents this speed as the number of queries that can be processed for 1 minute. In general,
the symbolic addresses processing slows down the solving process in several times. The last two columns display
the number of generated inputs (\texttt{SAT}) and how many of them are correct (\texttt{Accuracy}), i.e. actually
led to the inversion of targeted branch and have the same branch trace before it. We have only 3 programs,
that have been fully analyzed by both analyses: \texttt{libcbor}, \texttt{sqlite3}, and \texttt{yodl}.
The number of generated inputs for them shows that symbolic execution with addresses reasoning produces more inputs,
that is, allows to invert new additional branches. For other programs it have less number of SATs because the
default symbolic execution outruns it in selected time limit. The accuracy can't be equally
compared for the most of the programs, because of the different proportions of inverted branches. However,
in general the percentage of the correctly generated inputs remain at the same level. The only exception is
\texttt{libcbor} application, where symbolic execution accuracy has dropped from 100\% to 40.6\%. Even so,
according to the Table~\ref{tbl:sym_branches} the symbolic addresses reasoning allows to find new branches
for this application. Overall, the Table~\ref{tbl:analysis} shows that although the path predicate building
time is increased, the most critical part for symbolic execution is SMT-solving efficiency.

\begin{table}[ht]
\caption{Symbolic Branches}
\begin{center}
\scriptsize
\begin{tabular}{l @{\hspace{0.9\tabcolsep}} r >{\columncolor[gray]{0.9}}r r >{\columncolor[gray]{0.9}}r r >{\columncolor[gray]{0.9}}r}
\toprule
    \multirow{2}{*}{\textbf{Application}}&\multicolumn{2}{c}{\textbf{Total}}&\multicolumn{2}{c}{\textbf{Unique}}&\multicolumn{2}{c}{\textbf{New and unique}} \\
    &\textbf{default}&\textbf{symaddr}&\textbf{default}&\textbf{symaddr}&\textbf{default}&\textbf{symaddr} \\
    cjpeg&6992&30098&150&233&3&86 \\
    libxml2&9840&16423&452&531&0&79 \\
    readelf&19790&23009&924&937&0&13 \\
    libcbor&122&158&31&34&0&3 \\
    openssl&7561&7804&200&220&0&20 \\
    sqlite3&6979&9001&55&67&0&12 \\
    minigzip&8977&52861&23&68&0&45 \\
    hdp&28227&30620&431&460&2&31 \\
    yices-smt2&10462&23497&94&555&0&461 \\
    yodl&6676&6992&65&79&0&14 \\
    jasper&771811&1093902&97&107&0&10 \\
    %muraster&7102&7109&75&82&0&7 \\
\bottomrule
\end{tabular}
\label{tbl:sym_branches}
\end{center}
\end{table}

Table~\ref{tbl:sym_branches} presents the number of symbolic branches discovered during analysis.
The first column of the table shows the total number of discovered symbolic branches.
This column depicts how much the symbolic part of the program increased when symbolic addresses reasoning is enabled.
For some programs the number of symbolic branches increased in several times. The second column shows the
number of unique branches, i.e. the branches that are distinguished by its module name, source, and destination
addresses. The number of unique branches in orders of magnitude less due to the presence of loops and
code reuse in programs. The number of unique branches determine the real quantity of the new discovered code.
The last column contains the number of those unique branches that are new for the alternative analysis run.
That is, the \texttt{symaddr} subcolumn contains the number of symbolic branches that were discovered when symbolic
addresses reasoning was enabled and weren't found with default symbolic execution. For all programs we were able to
find new branches when enabling addresses processing, and for the most of these programs we didn't loose any branches
compared to the default analysis. However, we lost the couple of branches for some applications (\texttt{cjpeg}
and \texttt{hdp}).  Most likely there are some flaws in our symbolic execution implementation which led to the
concretization of some branches. However, this situation needs a further researching.

The most significant result of the new feature implementation is how it affects program coverage.
To evaluate this we launched symbolic execution on the programs with the same input (file with expected
format individually for each program) twice. We set 90 seconds timeout for the single SMT-query solving,
also the optimistic solutions~\cite{yun18} were enabled. The result of analysis is the corpus of new input
files, each representing the discovered execution path. We launched Sydr with static caching, that is if some
branch was successfully inverted, then it was excluded from the following queries to prevent duplicate
invertion. After this, a total basic block coverage was computed for every application. We use DynamoRIO
\texttt{drcov} tool and IDA Pro plugin Lighthouse~\cite{lighthouse} to compute coverage. It allows us to
represent coverage in percentage of total binary code size.
\begin{table}[ht]
\caption{Program Coverage}
\begin{center}
\scriptsize
\begin{tabular}{l r >{\columncolor[gray]{0.9}}r r >{\columncolor[gray]{0.9}}r r >{\columncolor[gray]{0.9}}r}
\toprule
    \multirow{2}{*}{\textbf{Application}}&\multicolumn{2}{c}{\textbf{Code coverage (\%)}}&\multicolumn{2}{c}{\textbf{Coverage diff (\%)}} \\
    %&\textbf{default}&\textbf{symaddr}&\textbf{default\textbackslash symaddr}&\textbf{symaddr\textbackslash default} \\
    &\textbf{default}&\textbf{symaddr}&\textbf{default$\setminus$symaddr}&\textbf{symaddr$\setminus$default} \\
    cjpeg&19.58&20.82&0&1.25 \\
    libxml2&7.8&9.6&0&1.8 \\
    readelf&16&15.8&0.8&0.6 \\
    libcbor&70.43&59.17&14.4&3.14 \\
    openssl&5.19&5.25&0.02&0.08 \\
    sqlite3&5.5&5.6&0&0.1 \\
    minigzip&29.69&31.14&0&1.45 \\
    hdp&9.5&9.2&0.34&0.04 \\
    hdp(libmfhdf)&13.95&14.83&0.45&1.33 \\
    hdp(libdf)&9.18&8.65&0.65&0.12 \\
    yices-smt2&2.23&2.33&0&0.1 \\
    yodl&28.25&29.17&0&0.92 \\
    jasper&9.94&10.07&0&0.13 \\
\bottomrule
\end{tabular}
\label{tbl:coverage}
\end{center}
\end{table}
The results of this evaluation are presented in the
Table~\ref{tbl:coverage}. The column \texttt{Code coverage} shows the total coverage for each program, which was
achieved by both analyses. The second column depicts the difference in coverage between these two Sydr runs.
The subcolumn \texttt{symaddr\textbackslash default} shows the unique basic block coverage that was discovered by
analysis with symbolic addresses reasoning and was not discovered during default symbolic execution and vice versa for the
subcolumn \texttt{default\textbackslash symaddr}. That is, the second column shows which symbolic execution was more
productive, depending on which subcolumn has a greater value.

The experiment results show that for the half of the programs the analysis with enabled symbolic
addresses reasoning was able to discover new program coverage without any losses. The increase in coverage
for all programs is from insignificant 0.4\% (for \texttt{hdp}) to 23\% (for \texttt{libxml2}).
But for the half of applications it turned out that both analyses discovered their own unique program coverage,
that is, both analyses explored different parts of the program. Besides, some applications (\texttt{readelf},
\texttt{libcbor}, \texttt{hdp}) have more coverage explored by the default symbolic execution. This happens due to
more complex SMT-queries when symbolic addresses reasoning is enabled. As a result, some SMT queries weren't
able to be solved for the given timeout, some queries were reasoned as unresolvable because of additional
address expressions.

These conducted experiments show that when analyzing real-world programs it is not always useful to have an address
reasoning enabled all the time. The most promising way is to combine analyses from two separate runs: default symbolic
execution allows to explore program comparatively fast, and during the second run with symbolic addresses processing
it discovers new parts of the program in addition to which have already been covered. It is possible to use branch caching mechanisms to prevent exploring the same parts of the program in two runs.

\section{Conclusion}
\label{conclusion}

We implemented the symbolic addresses processing on memory reads in our dynamic symbolic execution tool Sydr.
Different memory modeling techniques and symbolic address ranges reasoning were considered.
The addressable memory region is determined by analyzing symbolic address expression AST and utilizing SMT-solver.
If these methods fails, then memory region of constant length is selected. Symbolic memory reads are modeled
with linearization method based on the one proposed by Mayhem~\cite{cha12}. We enhance it by combining with
horizontal lines, nested ITE tree for large memory accesses, and considering symbolized memory values.
This approach was compared with full nested ITE tree and BST methods by utilizing several SMT-solvers.
We discovered that our linearization approach produces SMT-queries that are faster processed by solver than
those, produced by other methods. Finally, we analyzed the set of real-world programs with default symbolic
execution and with symbolic addresses reasoning enabled. Despite the fact that the new feature radically slows down
the analysis, it helps to find many new symbolic branches and discover new program coverage. However, in some
cases it could lead to some coverage losses. So, the optimal way to utilize the symbolic reads processing is performing
symbolic execution in separate runs in addition to each other.

\printbibliography

\end{document}